\documentstyle[preprint,aps]{revtex}

\font\svrm = cmr7

\def\pmb#1{\setbox0=\hbox{$#1$}%
\kern-.025em\copy0\kern-wd0
\kern.05em\copy0\kern-\wd0
\kern-.025em\raise.0433em\box0 }

\def\ttt#1{%
\setbox5=\hbox{$#1$}%
\setbox6=\hbox{\the\scriptfont2\char'030}%
\ifnum\wd5>\wd6{\vbox{\offinterlineskip
\hbox to\wd5{\hfil\the\scriptfont2\char'030\hfil}%
\hbox to\wd5{\hfil\the\scriptfont2\char'030\hfil}%
\vskip1.4pt\hbox{$#1$}}}%
\else{\vtop{\offinterlineskip
\hbox{\the\scriptfont2\char'030}
\hbox{\the\scriptfont2\char'030}
\vskip1.4pt\hbox to\wd6{\hfil$#1$\hfil}}}\fi}

\def\be{\begin{equation}}
\def\ee{\end{equation}}
\def\beq{\begin{eqnarray}}
\def\eeq{\end{eqnarray}}
\def\diag{{\rm diag}}
\def\half{{\textstyle\frac{1}{2}}}

\def\square{\kern1pt\vbox{\hrule height1.2pt\hbox{\vrule width1.2pt\hskip3pt
\vbox{\vskip 6pt}\hskip3pt\vrule width0.6pt}\hrule height0.6pt}\kern1pt}

\begin{document}

\def\footnoterule{\hrule width \hsize}
\skip\footins = 20pt
\footskip     = 20pt
\footnotesep  = 20pt


\textwidth=6.75in
\hsize=6.9in
\oddsidemargin=0in
\evensidemargin=0in
\hoffset=-.2in

\textheight=9.4in
\vsize=9.4in
\topmargin=-.5in
\voffset=-.3in

\setcounter{page}{0}

\title{Quantal Analysis of String-Inspired Lineal Gravity with~Matter~Fields%
\footnotemark[1]}

\footnotetext[1]
{\advance\baselineskip by -8pt
This work is supported in part
by funds provided by the U.S.~Department of Energy (D.O.E.)
under contract \#DE-FC02-94ER40818 (RJ),
by the U.S.\ National Science Foundation (N.S.F.) under contract
\#PHY-89-15286
and by the Swiss National Science Foundation (DC).}

\author{D. Cangemi}
\medskip
\address{Physics Department, University of California at Los Angeles\\
405 Hilgard Ave., Los Angeles, CA ~90024--1547}

\bigskip

\author{R. Jackiw}
\medskip

\address{Center for Theoretical Physics, Laboratory for Nuclear Science
and Department of Physics \\
Massachusetts Institute of Technology, Cambridge, MA ~02139--4307}


\maketitle

\begin{abstract}

We show that string-inspired lineal gravity interacting with matter fields
cannot be Dirac-quantized owing to the well known anomaly in
energy-momentum tensor commutators.

\end{abstract}

\thispagestyle{empty}

\widetext

\vspace{\fill}

\begin{center}
Submitted to: {\em Physics Letters B}
\end{center}
\vskip36pt
\noindent
\hbox to \hsize{MIT-CTP-2300, UCLA/94/TEP/18
\hfil hep-ph/9405119 \hfil May 1994}
\vskip-12pt

\newpage

Since the string-inspired model for lineal gravity with matter fields has been
extensively studied within various semi-classical approximations,%
\footnote{For a summary, see \cite{ref1}.}
we are led to inquire
whether the quantized theory can be analyzed exactly.  Quantal
results for this model without matter \cite{ref2,ref3,ref4}
or with point particles \cite{ref4} are now
available.  Here we consider this gravity theory in the presence of a massless
scalar field.

The model is formulated as a gauge theory \cite{ref5}
of the extended Poincar\'e group \cite{ref6} and
the analysis is carried out in a Schr\"odinger representation, so that issues
related to functional integration do not arise.  Our results are the
following.

\begin{itemize}
\item[1.]
The quantum theory can be presented in a well-defined, unambiguous fashion,
thus putting to rest the worry that ``there is not a unique quantization of
dilaton gravity''~\cite{ref1}.
Evidently the gauge principle resolves ambiguities.

\item[2.]
All dynamical variables can be separated and decoupled.
The operator equations of motion can be solved.

\item[3.]

There is no interaction between matter and gravity degrees of freedom, save a
correlation interaction induced by diffeomorphism constraints.

\item[4.]
The diffeomorphism constraints cannot be satisfied owing to a commutator
anomaly, so a diffeomorphism invariant
state space for the quantum theory cannot be constructed.

\item[5.]
A semi-classical reduction of the fully quantized model fails to reproduce
familiar results.

\end{itemize}

\noindent
Detailed exposition of our investigation
will be presented elsewhere \cite{ref7};
here we give a brief description.

The gauge group is the 4-parameter extended Poincar\'e group.
The Lie algebra possesses a central element $I$
in the commutator of translations $P_a$;
the Lorentz generator $J$ satisfies
conventional commutation relations.

\be
{}[P_a,P_b] = \epsilon_{ab} I ~~,~~~~ [P_a,J] = \epsilon_a^{~b} P_b
\label{eq1}
\ee
[Notation: tangent space is indexed by $(a,b,\ldots)$ with metric tensor
$h_{ab} =
\diag (1,-1)$ and $\epsilon^{ab} = -\epsilon^{ba}$, $\epsilon^{01} = 1$.  Light
cone components will be frequently employed; they are defined by $(\pm)\equiv
{1\over\sqrt{2}} \Big( (0) \pm (1) \Big)$.]  The gauge connection $A_\mu$ is
identified with metric quantities,
\be
A_\mu = e_\mu^a P_a + \omega_\mu   J + a_\mu I
\label{eq2}
\ee
where $e_\mu^a$ is the {\it Zweibein\/}, determining      the metric tensor
        of space-time, $g_{\mu\nu} = e_\mu^a e_\nu^b h_{ab}$;
$\omega_\mu$ is the spin-connection --- an independent    variable     at the
initial    stage;   $a_\mu$ is a $U(1)$ gauge potential     associated    with
the     central    element,      which gives rise          to a cosmological
constant.         A gauge   group       transformation       $U$,
parameterized   as $U = e^{\theta^a P_a} e^{\alpha J} e^{\beta I}$,     acts
on the      connections      by (inhomogeneous) adjoint action,
\begin{mathletters}%
\label{eq3}
\beq
e_\mu^a &\to&  \left( e^U \right)_\mu^a = \left( \Lambda^{-1} \right)^a_{~b}
\left( e_\mu^b + \epsilon^b_{~c} \theta^c \omega_\mu +
\partial_\mu \theta^b \right)
\label{eq3a}\\
\omega_\mu &\to&  \left( \omega^U \right)_\mu = \omega_\mu + \partial_\mu
\alpha
\label{eq3b}\\
a_\mu &\to&     \left( a^U \right)_\mu = a_\mu - \theta^a \epsilon_{ab}
e_\mu^b -
\half \theta^a \theta_a \omega_\mu + \half \partial_\mu \theta^a \epsilon_{ab}
\theta^b
\label{eq3c}
\eeq
\end{mathletters}%
where $\Lambda^a_{~b}$ is the Lorentz transformation matrix:
 $\Lambda^a_{~b} = \delta^a_{~b} \cosh \alpha +
\epsilon^a_{~b} \sinh \alpha$.
In the usual    way,     one constructs     from (\ref{eq1}) and (\ref{eq2})
the gauge group curvature.
\beq
F \equiv \half \epsilon^{\mu\nu} F_{\mu\nu} &=& \epsilon^{\mu\nu}
\left(
\partial_\mu A_\nu + A_\mu A_\nu \right)
= f^a P_a + f^2 J_2 + f^3 I \nonumber\\
&=& \epsilon^{\mu\nu} \biggl\lbrace
\left(
\partial_\mu e_\nu^a +
\epsilon^a_{~b} \omega_\mu e_\nu^b \right) P_a +
\partial_\mu \omega_\nu J + \left(
\partial_\mu a_\nu + \half e_\mu^a \epsilon_{ab} e_\nu^b \right)
I \biggr\rbrace
\label{eq4}
\eeq

To build a gauge and diffeomorphism     invariant       action,
a Lagrange multiplier multiplet    $(\eta_a, \, \eta_2, \, \eta_3)$ is
introduced,     and is taken     to transform       in the     coadjoint
representation.\footnote{\advance\baselineskip by -8pt
Even though the group is not semi-simple, one can
define an invariant, non-singular metric, so the
coadjoint action can be identified with the adjoint; see Ref.~6.}
\begin{mathletters}
\label{eq5}
\beq
\eta_a  &\to& \left( \eta^U \right)_a  = \left( \eta_b -
\eta_3 \epsilon_{bc} \theta^c \right) \Lambda^b_{~a} \label{eq5a}\\
\eta_2  &\to& \left( \eta^U \right)_2  = \eta_2 - \eta_a \epsilon^a_{~b}
\theta^b - \half \eta_3 \theta^a \theta_a \label{eq5b}\\
\eta_3  &\to& \left( \eta^U \right)_3  = \eta_3 \label{eq5c}
\eeq
\end{mathletters}%
The gauge invariant gravitational    action   then reads
\be
I_g =
{1\over4\pi G}
\int d^2x ~ \left( \eta_a f^a + \eta_2 f^2 + \eta_3 f^3 \right)
\label{eq6}
\ee
where $G$ is ``Newton's constant'', which is dimensionless, as are the
Lagrange multiplier fields.  (The velocity of light and $\hbar$ are scaled to
unity.)  It is straightforward to show that $I_g$ is equivalent to the
action ${1\over4 \pi G} \int d^2 x \sqrt{-g} \left( e^{-2\bar{\varphi}} R -
\lambda
\right) $ where $e^{-2\bar{\varphi}}$ is proportional to $\eta_2$ with the
cosmological constant $\lambda$ arising as the solution to the equation for
$\eta_3$ that follows from~(\ref{eq6}).  A further redefinition
$\bar{g}_{\mu\nu} = e^{2\bar{\varphi}} g_{\mu\nu}$
relates (\ref{eq6}) to the string-inspired ``dilation'' action
${1\over4\pi G}  \int d^2x \sqrt{-\bar{g}} e^{-2\bar{\varphi}} \left(
\bar{R} + 4 \bar{g}^{\mu\nu} \partial_\mu \bar{\varphi} \partial_\nu
\bar{\varphi} - \lambda \right)$.

A diffeomorphism    invariant        action       for   a
scalar field $\varphi$ with mass $\mu$ is of course the familiar expression
$I_m^{\,0} = {1\over2} \int d^2 x e
\left(
E_a^\mu E_b^\nu h^{ab}
\partial_\mu \varphi \partial_\nu \varphi - \mu^2 \varphi^2
\right) ~,~~ e \equiv \det e^a_\mu, ~
E_a^\mu e_\mu^b = \delta^a_b$.
However, with $\varphi$ taken to be gauge invariant,
$I_m^{\,0}$
is not invariant against local (gauge) translations
[see (\ref{eq3a})].

Construction of a Poincar\'e    gauge     invariant matter
action     makes use             of a new field    $q^a$ ---    the
``Poincar\'e
 coordinate'' ---  and its       canonical        conjugate $p_a$.     Both
transform     as tangent-space
Lorentz    vectors;     additionally   $q^a$   shifts     by
a translation.
\begin{mathletters}
\beq
q^a &\to&   (q^U)^a =    (\Lambda^{-1})^a_{~b}
\left( q^b + \epsilon^b_{~c} \theta^c \right)
\label{eq8a} \\
p_a &\to& (p^U)_a = p_b \Lambda^b_{~a}
\label{eq8b}
\eeq
\end{mathletters}%
As a consequence    $q^a$   can be set     to zero    by a gauge
transformation.

One verifies      that the     following      Lagrange density    for
matter   variables is invariant against transformations (3) and (7).
($\Pi$, $\varphi$, and $u,\,v$ are not transformed.)
\beq
{\cal L}_m &=& p_a \dot{q}^a + \Pi \dot{\varphi} + e_0^a
 \epsilon_a^{~b} p_b  + \omega_0
 p_a \epsilon^a_{~b} q^b  - u {\cal E} - v {\cal P}
\label{eq9}
\eeq
The Lagrange multipliers $u$ and $v$ enforce the vanishing of the energy
density ${\cal E}$ and momentum density ${\cal P}$ --- these are the
diffeomorphism constraints.
\begin{mathletters}
\beq
{\cal E} &\equiv&
(Dq)^a \epsilon_a^{~b} p_b
+ \half \left( \Pi^2 + (\varphi')^2 + (Dq)^2 \mu^2 \varphi^2 \right)
\label{eq10a}\\
{\cal P} &\equiv& p_a \left( D q \right)^a + \Pi \varphi'
\label{eq10b}
\eeq
\end{mathletters}%
Here
$(Dq)^a \equiv q'^a + \epsilon^a_{~b} (q^b \omega_1 - e_1^b)$,
while dot/dash denote time/space differentiation.\footnote
{\advance\baselineskip by -8pt
Our construction of the gauge invariant matter action follows the method given
by G.~Grignani and G.~Nardelli (Ref.~8).   The haphazard-appearing
expressions (8), (9) can    in fact    be systematically derived within an
explicitly gauge covariant formalism, see Ref.~\cite{ref7}.\hfil\break}

The $\varphi$-field dynamics implied by (8) and (9) are the same as those
arising from $I_m^{\,0}$. This is seen by passing to the gauge $q^a = 0$ and
eliminating $p_a$ from (8) by setting ${\cal E}$
and ${\cal P}$ to zero.  Gauge invariance has been achieved by a
Higgs-like mechanism, and vanishing $q^a$ corresponds to a ``unitary gauge''
wherein the physical content of a gauge-invariant Lagrangian is exposed.

The complete Lagrange density for the gravity-matter system
 that we study follows from (6), (8) and (9).\footnote
{\advance\baselineskip by -8pt
In two dimensions, non-minimal coupling to the gravitational gauge potentials
is also possible, see Ref.~[6], but we do not make use of it here.}
\beq
{\cal L}_{g+m} &=& {1 \over 4\pi G}
\left( \eta_a \dot{e}_1^a
+ \eta_2 \dot{\omega}_1
+ \eta_3 \dot{a}_1 \right)
+ p_a \dot{q}^a + \Pi \dot{\varphi}
+ e_0^a G_a + \omega_0 G_2 + a_0  G_3    -     u {\cal E} - v {\cal P}
\label{eq11}
\eeq
(Spatial, but not temporal, integration       by parts is
carried out freely.)   The symplectic     structure     identifies   the
canonical     coordinates    as $(e_1^a, \omega_1, a_1), ~ q^a$
and $\varphi$, while their     respective     canonical momenta are
${1\over 4 \pi G} \left( \eta_a , \eta_2, \eta_3 \right),~ p_a$
and $\Pi$.   The Hamiltonian is a superposition     of
the diffeomorphism constraints and the gauge constraints;
$(e_0^a, \omega_0, a_0)$ enforce the vanishing of the gauge generators
$(G_a, G_2, G_3)$.
\begin{mathletters}%
\beq
G_a &=& {1\over 4 \pi G}
\left(
\eta'_a + \epsilon_a^{~b} \eta_b \omega_1 + \eta_3 \epsilon_{ab} e_1^b
\right)
+ \epsilon_a^{~b} p_b
\label{eq12a} \\
G_2 &=& {1\over 4\pi G}
\left( \eta'_2 + \eta_a \epsilon^a_{~b} e_1^b \right)
- q^a \epsilon_a^{~b} p_b
\label{eq12b} \\
G_3 &=& {1\over 4 \pi G} \eta'_3
\label{eq12c}
\eeq
\end{mathletters}%

Using the Poisson brackets implied by the symplectic structure, one verifies
that the algebra      of constraints      closes;    they are first-class.
The gauge generators  commute with ${\cal E}$ and ${\cal P}$,
while among themselves they follow the Lie algebra.
\begin{mathletters}
\beq
\left[ G_a (x), G_b (y) \right]_{PB} = \epsilon_{ab} G_3 (x)
{} ~ \delta (x-y)
\label{eq13a} \\
\left[ G_a (x), G_2 (y) \right]_{PB} = \epsilon_a^{~b} G_b (x)
{} ~ \delta (x-y)
\label{eq13b}
\eeq
\end{mathletters}%
The diffeomorphism constraints satisfy
\begin{mathletters}
\beq
\left[ {\cal E} (x), {\cal E} (y) \right]_{PB}
&=& \left[ {\cal P}(x), {\cal P}(y) \right]_{PB}
= \left( {\cal P} (x) + {\cal P} (y) \right) \, \delta' (x-y)
\label{eq14a} \\
\left[ {\cal E} (x) , {\cal P} (y) \right]_{PB}
&=&
\left( {\cal E}(x) + {\cal E}(y) \right) \,
\delta' (x-y) ~~.
\label{eq14b}
\eeq
Eqs.~(13a,b) may be decoupled by defining
$T_\pm = \half ({\cal E} \pm {\cal P})$
\beq
\left[ T_\pm (x) , T_\pm (y) \right]_{PB} &=&
\pm \left( T_{\pm}(x) + T_{\pm}(y) \right) \, \delta'(x-y)
\label{eq15a} \\
\left[ T_\pm(x) , T_{\mp}(x) \right]_{PB} &=& 0
\label{eq15b}
\eeq
\end{mathletters}%
[A common time argument in the above quantities has been suppressed.]  Note
that (13) coincides with the algebra of a conformally invariant theory, even
though the field is massive.

Quantization      consists    of replacing  Poisson brackets by
commutators.  The question   of the quantum nature    of the above
constraint algebra   will not be   addressed as yet.       Rather    we set
ourselves the task of satisfying       the requirement     of vanishing
constraints
by attempting to solve the corresponding
functional differential equations that a quantum mechanical         wave
functional    $\Psi$,    in the Schr\"odinger       representation,
must satisfy.
Moreover, we do not well-order operators in the constraints at intermediate
steps of the calculation; the ordering is stipulated only at the end, when a
constraint is taken to act on the wave functional.
We work            in ``momentum'' space     for the   metric
variables   and in ``position'' space   for the matter  variables {\it i.e.},
$\Psi$     depends    on $(\eta_a, \eta_2, \eta_3)$, $q^a$ and $\varphi$ while
$(e_1^a,~\omega_1,~a_1)$ are realized as the functional derivatives
$4\pi G i \left(
{\partial \over \partial \eta_a},\,
{\partial \over \partial \eta_2},\,
{\partial \over \partial \eta_3} \right)$
and similarly $p_a$ and $\Pi$ as
${1\over i} \, {\delta \over \delta q^a}$ and
${1\over i} \, {\delta \over \delta \varphi}$.

Before proceeding with the     task of solving      the
quantum constraints, we record the
general classical
solution to the equations that   follow from (9), (10), (11).  Henceforth, to
simplify
the problem     and to make      contact with previous work, we set $\mu$, the
mass
of $\varphi$, to zero.  The general solution to the equations of motion is the
following.

The equations for the metric variables do not involve the matter variables;
they require the vanishing of $F$, hence $A_\mu$ is a pure gauge and
explicitly is given by
\begin{mathletters}
\beq
e_\mu^\pm &=& \exp \left( \pm \alpha \right) \,  \partial_\mu \theta^\pm
\label{eq16a} \\
\omega_\mu &=& \partial_\mu \alpha
\label{eq16b} \\
a_\mu &=& \partial_\mu \beta + \half \epsilon_{ab} \partial_\mu \theta^a
\theta^b
\label{eq16c}
\eeq
\end{mathletters}%
where $\theta^a$, $\alpha$, $\beta$ are arbitrary functions
of space-time $(t,x)$ and specify an arbitrary gauge transformation.

To solve the matter-field equations, we choose    two functions of space-time,
$X^a (t,x)$, and parameterize $u$ and $v$ as
\be
v \pm u = \dot{X}^\pm / X'^\pm
\label{eq17}
\ee
Also we introduce six arbitrary mode functions of the single variable $X^+$ or
$X^-$: $N_\pm (X^\pm)$, $Q^\pm (X^\pm)$, $\phi_\pm (X^\pm)$.  The general
solution, which satisfies all the equations, save the diffeomorphism
constraints, may be presented as:
\begin{mathletters}
\beq
q^\pm &=& e^{\pm\alpha}  \left( Q^\pm (X^\pm) \mp \theta^\pm \right)
\label{eq17a} \\
p_{\pm} &=& \mp e^{\mp\alpha}  {\partial \over \partial x}
N_\pm (X^\pm)
\label{eq17b}
\eeq
\end{mathletters}%
\begin{mathletters}
\beq
\eta_3 &=& \lambda
\label{eq18a} \\
{1\over 4\pi G} \eta_{\pm} &=& e^{\mp \alpha}
\left( N_\pm (X^\pm) \mp {\lambda\over 4\pi G} \theta^\mp \right)
\label{eq18b} \\
{1\over 4\pi G} \eta_2 &=&
- \int^{X^+} dz ~ N'_+ (z) Q^+ (z)
- \int^{X^-} dz ~ N'_- (z) Q^- (z)
\nonumber\\
&& \hbox{\qquad} + N_+(X^+) \theta^+ - N_- (X^-) \theta^-
- {\lambda\over 4\pi G} \, \theta^+ \theta^-
\label{eq18c}
\eeq
\end{mathletters}%
\begin{mathletters}
\beq
\varphi &=& \phi_+ (X^+) + \phi_- (X^-)
\label{eq19a} \\
\Pi &=& {\partial \over \partial x}
\, \left( \phi_+ (X^+) - \phi_-(X^-) \right)
\label{eq19b}
\eeq
\end{mathletters}%
Finally the diffeomorphism constraints that ${\cal E}$ and ${\cal P}$ vanish
are satisfied by identifying $\phi_\pm$ with $N_\pm$, $Q_\pm$,
\be
\left( \phi'_\pm \right)^2 = N'_\pm {Q'^\pm}
\label{eq22}
\ee
[In (17) and (19) the dash on the mode functions signifies differentiation
with respect to argument; {\it viz.\/}
${\partial \over   \partial x}
\phi_\pm (X^\pm) = X'^\pm \phi'_\pm (X^\pm)$, {\it etc\/}.]

The form of the general   solution illustrates well the flexibility inherent
in a gauge theory of gravity.     With        all functions arbitrary, neither
the geometry    nor the field motion are specified; indeed upon setting the
gauge parameters $(\theta^a, \alpha, \beta)$ to zero,
$(e_\mu^a, \omega_\mu, a_\mu)$ vanish   and    no geometry   can be
constructed.    The ``unitary gauge'' of vanishing $q^a$ is achieved by
choosing $\theta^a = -\epsilon^a_{~b} Q^b$.
Metric variables are no longer zero; they describe vanishing curvature,
without selecting a specific coordinate system on which the embedding
functions $X^a$ are defined.   To select coordinates it is natural to set
$e_\mu^a = \delta_\mu^a$, $\omega_\mu = 0$,
{\it i.e.\/} $\theta^a = x^a,~ \alpha = 0$
and also $\beta$ is taken to vanish so that    $a_\mu = \half
\epsilon_{\mu\nu} x^\nu$.
It follows that $X^\pm$ depends only on $x^\pm$ and coincides with the
function inverse to $Q^\pm$, but its specific form is still open.   A simple
choice  is $X^a = x^a$, so that $u=1,~v=0$.
Eq.~(18) then shows that the matter field
$\varphi$ consists of a superposition of left-moving and right-moving waves,
$\phi_\pm$, while
(19) expresses $N_\pm$ --- the remaining mode function --- in terms of
$\phi_\pm$.

We now return to the problem of solving the quantum mechanical theory; {\it
viz.\/} solving the gauge and diffeomorphism constraints.   We begin by
recording         the known solution      in the pure gravity  case, with
the matter variables $(p_a, q^a, \Pi, \varphi)$ omitted.  Only the gauge
constraints
act, and they are solved by
\be
\Psi(\eta) \bigg|_{\hbox{\svrm pure gravity}}
= \delta (\eta'_3 ) \, \delta (M') ~ e^{i\Omega}
\Psi(M,\lambda) \biggl|_{\hbox{\svrm pure gravity}}
\label{eq21}
\ee
Here $M$ is the gauge-invariant combination
\be
M = \eta_a \eta^a - 2 \eta_2 \eta_3
\ee
and $\Omega$ is the Kirillov-Kostant 1-form on the coadjoint orbit of the
extended Poincar\'e group.
\be
\Omega = {1\over 8 \pi G \, \lambda} \int \epsilon^{ab} \eta_a d \eta_b
\label{eq24}
\ee
The pure gravity constraints require that the wave functional depends only on
the constant parts of $\eta_3$ and $M$; we call the former $\lambda$.
(In the classical theory, the constant
$M$ corresponds to the ``black hole'' mass,
and $\lambda$ to the cosmological constant.)

In the presence of the matter field, $G_3$ still requires that $\eta_3$ be the
constant $\lambda$.
It is then convenient to separate the Kirillov-Kostant
phase factor from the wave functional.  Furthermore, $q^a$ is shifted by
$\eta^a / \lambda$, and the new variable
\be
\rho^a \equiv q^a + \eta^a / \lambda
\label{eq25}
\ee
responds only to Lorentz gauge transformations, while it is translation
and $U(1)$ invariant.
This is identical to the transformation law for $p_a$, which is now taken as
conjugate to $\rho^a$.
Also $\eta_2$ is shifted by $\eta_a \eta^a / 2
\lambda$, so that $-2\lambda\eta_2$ is replaced by
the gauge invariant variable $M$.
Finally,
${1\over4\pi G}\omega_1$, the coordinate conjugate to $\eta_2$, is renamed
$2\lambda\Pi_M$.
With
these redefinitions, the $G_a$ constraint requires that there be no further
dependence on $\eta_a$.  Thus the wave-functional satisfying $G_3$ and $G_a$
takes the form
\be
\Psi = \delta({\eta_3}') \, e^{i \Omega}
\widetilde{\Psi} (\rho^a , M , \varphi)
\label{eq26}
\ee
and the remaining gauge constraint $\widetilde{G}_2$
in terms of new variables is
\beq
-\widetilde{G}_2 &=& {1\over8\pi G \lambda} M' + \rho^a \epsilon_a^{~b} p_b
\label{eq26b}
\eeq
while the diffeomorphism constraints read,
\begin{mathletters}
\beq
\widetilde{\cal E} &=& \rho'^a \epsilon_a^{~b} p_b -
8 \pi G \lambda \Pi_M \rho^a p_a - {4\pi G \over \lambda}
p_a p^a + \half \left( \Pi^2 + \varphi'^{2} \right)
\label{eq27a}\\
\widetilde{\cal P} &=& p_a \rho'^a  - 8 \pi G \lambda \Pi_M \rho^a
\epsilon_a^{~b} p_b +  \Pi\varphi'
\label{eq27b}
\eeq
\end{mathletters}%
Effectively $\widetilde{\Psi}$ is governed by the Lagrange density
\be
\widetilde{\cal L}_{g+m} = p_a \dot{\rho}^a + \Pi_M\dot{M} + \Pi \dot{\varphi}
+ \omega_0 \widetilde{G}_2 - u {\widetilde{\cal E} - v \widetilde{\cal P}} ~~.
\label{eq29}
\ee
One verifies  that $\widetilde{\cal L}_{g+m}$
is obtained from ${\cal L}_{g+m}$ in (10) by adding a
total time derivative, solving the $G_a$ and $G_3$ constraints,
and redefining variables as indicated above.

To unravel the $\widetilde{G}_2$ constraint,     we extract     from
the wave functional       an $M$-dependent phase factor.
Also we shift the $M$ dependence in the remaining functional
by a term proportional to $\rho^a \rho_a$, which is equivalent to a canonical
transformation from $M$   and $p_a$ to $m = M - {\lambda^2 \over 2} \rho^a
\rho_a$ and $\Pi_a = p_a + {\lambda^2} \rho_a \Pi_M$
($\rho^a$ and $\Pi_M = \Pi_m$ are unaffected).
\beq
\widetilde{\Psi} (\rho^a, M, \varphi)
&=& e^{i\widetilde{\Omega}} \ttt{\Psi}
(\rho^a,m,\varphi)
\label{eq30}\\
\widetilde{\Omega} &=& {1\over 8 \pi G\lambda} \int d \hat{\rho}^a
\epsilon_{ab}
\hat{\rho}^b M
\label{eq31}\\
\hat{\rho}^a &\equiv& \rho^a / \rho ~~,~~~~ \rho \equiv \sqrt{\rho^a \rho_a}
\nonumber
\eeq
The constraints on $\ttt{\Psi}$ now become
\be
-\ttt{G}_2 = \rho^a \epsilon_a^{~b} \, \Pi_b \equiv j
\label{eq32a}
\ee
\begin{mathletters}
\beq
\ttt{\cal E} &=&
- { 4 \pi G \over \lambda} \Pi_\rho^2
- {\lambda\over16\pi G} \rho'^2
+ 4 \pi G \lambda^3 \Pi_m^{\,2} \rho^2 +
{(m' - 8 \pi G \lambda j)^2 \over 16 \pi G \lambda^3 \rho^2}
+ \half \left( \Pi^2 + \varphi'^{2} \right)
\label{eq33a}\\
\ttt{\cal P} &=&  \Pi_\rho \rho'
+ \Pi_m (m' - 8\pi G \lambda j) + \Pi \varphi'
\label{eq33b}
\eeq
\end{mathletters}%
where $\Pi_\rho = \hat{\rho}^a \Pi_a$.
The $\ttt{G}_2$ constraint now simply requires
that $\ttt{\Psi}$
depends on $\rho^a$    only through its   magnitude $\rho$, and $j$
disappears from $\ttt{\cal E}$ and $\ttt{\cal P}$.

In the final step, we present $\ttt{\Psi}$ as a Fourier
transform with respect to $m'$, and call the conjugate variable
${1\over 8\pi G \lambda} \gamma$.
(This leaves the constant part of $m$ undetermined.)
\be
\ttt{\Psi} (\rho^a, m, \varphi)
= \int {\cal D} \gamma
\exp \left( {{i \over 8\pi G \lambda} \int m \, d \gamma} \right) \,
{\Phi} (\rho, \gamma, \varphi)
\label{eq34}
\ee
Acting on ${\Phi}$,
$\ttt{\cal E}$ and $\ttt{\cal P}$ become
\message{Print Me First!}
\begin{mathletters}
\beq
{\cal E} &=&
-{4\pi G \over \lambda}
\left( \Pi_\rho^2 - {1\over \rho^2} \Pi_\gamma^2 \right)
- {\lambda \over 16 \pi G} \left( \rho'^2 - \rho^2 \gamma'^2 \right)
+ \half (\Pi^2 + \varphi'^2)
\label{eq35a}\\
{\cal P} &=& \Pi_\rho \rho' + \Pi_\gamma \gamma' + \Pi\varphi'
\label{eq35b}
\eeq
\end{mathletters}%
where $\Pi_\gamma = {1\over i} \, {\delta \over \delta \gamma}$.
But now we see that the variables $(\rho,\gamma)$ and $(\Pi_\rho, \Pi_\gamma)$
can be interpreted     as the radial    and ``angular''     coordinates
according to the decomposition
\be
r^a = (\rho\cosh\gamma, \rho\sinh\gamma)
\label{eq36}
\ee

Thus the final effective Lagrange density is
\beq
{\cal L} &=&  \pi_a \dot{r}^a + \Pi \dot{\varphi}  - u {\cal E} - v {\cal P}
\label{eq37}
\eeq
with
\begin{mathletters}
\beq
{\cal E} &=&  - {1 \over 2}
\left( {g \pi^a \pi_a} + {1\over g} r'^a r'_a \right)
+ \half \left( \Pi^2 + \varphi'^2 \right)
{} ~~~~~~~ g = 8 \pi G / \lambda
\label{eq38a}\\
{\cal P} &=& \pi_a r'^a + \Pi \varphi' ~~.
\label{38b}
\eeq
\end{mathletters}%
The wave functional $\Phi(r^a,\varphi)$
must satisfy the energy and momentum constraints
which are completely decoupled and separated,
and there is no matter-gravity interaction.\footnote
{\advance\baselineskip by -8pt
Absence of matter-gravity interactions in the string-inspired model has also
been claimed in~\cite{ref9}.  Also we have shown \cite{ref4} that point
particles do not experience gravitational forces.}
Nevertheless the constraints give rise to
correlations between the variables, which are otherwise
non-interacting.

In fact the final formulas (35), (36)
may be gotten by another, shorter route.   In the
above approach, we have not chosen       a gauge, but arrived   at the final
expressions by solving the gauge constraints.  Alternatively, we may return
to (25)--(28)
and fix the Lorentz gauge freedom by setting
${1\over 8\pi G \lambda} \omega_1 = \Pi_M$ to
zero.
Since the bracket $[G_2, \Pi_M]$ is a c-number,
no compensating terms are needed.
The effective Lagrangian in this gauge reads
\beq
\widetilde{\cal L}_{g+m} \biggl|_{\Pi_M = 0}
 &=& p_a \dot{\rho}^a + \Pi \dot{\varphi}
+ \omega_0 \widetilde{G}_2  - \left( u \widetilde{\cal E} + v \widetilde{\cal
P} \right)
\biggl|_{\Pi_M=0}
\label{eq39}
\eeq
\begin{mathletters}
\beq
\widetilde{\cal E}
\biggl|_{\Pi_M=0}
&=& \rho'^a \epsilon_a^{~b} p_b -
{4\pi G\over \lambda} p^2
+ \half \left( \Pi^2 + \varphi'^2 \right)
\label{eq40a} \\
\widetilde{\cal P}
\biggl|_{\Pi_M=0}
&=& p_a \rho'^a + \Pi \varphi'
\label{eq40b}
\eeq
\end{mathletters}%
and with the redefinitions
\be
\pi_a = p_a + {\lambda \over 8\pi G} \epsilon_{ab} \rho'^b
{} ~~,~~~~ r^a = \rho^a
\label{eq41}
\ee
we again arrive at (35), (36),
with the two Lagrangians differing by a total time derivative
(equivalent to factoring from the wave
functional the phase
$e^{i {\lambda \over 16\pi G} \int \rho^a \, \epsilon_{ab} \, d\rho^b}$).
Also in (37), there still remains the $\widetilde{G}_2$
constraint (25), now transformed,
which requires
 that
$\left( M - {\lambda^2\over2} \rho^a \rho_a \right)'
+ j$ vanishes; but this merely serves to specify the $M$-dependence of the
wave-functional, and does not  interfere with the diffeomorphism
constraints. We now proceed to analyze them.

The momentum constraint in (36b), enforced by $v$, is easily
satisfied.   It requires that the wave functional  be invariant against
arbitrary reparametrization of the line coordinate $x$, with the fields
$\varphi,\,r^a$ transforming as scalars.  This is achieved when the dependence
on the fields is contained in integrals of 1-forms, {\it e.g.}
$\int {\cal F} (\varphi, r^a) \varphi' dx
= \int {\cal F} (\varphi,\,r^a) d\varphi$.

But we need not concern ourselves with satisfying this constraint: the
momentum density arises when energy densities are commuted [see (13a)], so we
expect that once the energy constraint is satisfied, the momentum constraint
will also be met.

However, the remaining energy constraint cannot be solved for the following
reason.  The energy density consists of three identical free-field terms
(for $r^0$, $r^1$, and $\varphi$) with the two gravitational  contributions
entering with opposite signs, regardless of sign ($8\pi G/\lambda$),
while the matter contribution is positive.
[The reversal of signs in the gravity terms is also
understood from a degrees-of-freedom count: metric gravity in two space-time
dimensions has a single negative degree of freedom, while the dilaton carries
a positive degree of freedom.]  The three fields form an $O(2,1)$
invariant array.

It is well known that the quantum commutator of energy density with momentum
density for free fields possesses an anomalous, non-canonical c-number
contribution proportional to $\delta'''(x-y)$.  This Schwinger term cancels in
the gravitational  contribution, owing to the opposite signs in the energy
densities of $r^0$ and $r^1$.  But the matter contribution survives,
producing an obstruction to solving the energy constraint:  the corresponding
(functional) differential equation is not integrable.

Various proofs have been given for the necessary presence of a triple
derivative Schwinger term in the energy-density -- momentum-density
commutator.  One approach uses    energy-momentum  conservation and
positivity of the Hilbert space.   Alternatively one may normal order the
field bilinears with respect to a pre-selected Fock vacuum.    Perhaps neither
argument is appropriate in a gravity context, where concepts of energy and
momentum are elusive and the vacuum state may be very unconventional, with no
relation to a normalizable Fock vacuum.  It is therefore fortunate, for our
argument, that an alternative proof can be devised, which makes no reference
to a vacuum state, but relies on manipulations in the functional space of the
Schr\"odinger representation.  The argument, which has been presented
elsewhere \cite{ref10}, proceeds as follows.

Consider a canonical free field $(\Pi, \varphi)$ and form the combination
$\chi = {1\over\sqrt{2}} (\Pi + \varphi')$.
The sum of the energy and momentum densities is given by $\chi^2$ and one
wants to determine the commutator of $\chi^2(x)$ with $\chi^2(y)$;
or better, the commutator $T_{f_1}$ with $T_{f_2}$ where
\be
T_f = \half \int dx \, \chi(x) \, f(x) \, \chi(x)
\label{eq42}
\ee
Owing to the necessarily singular nature of the operator product of $\chi$
with itself, (40) requires regularization and subtraction, so that the
regulated and subtracted quantity possesses a limit as the regulator is
removed, but an anomaly in the commutator may emerge.  To regulate, we replace
$T_f$ by $T_F$
\be
T_F = \half \int dx \, dy \, \chi(x) \, F(x,y) \, \chi(y)
\label{eq43}
\ee
and inquire what should be subtracted from $T_F$ so that the limit
\be
F(x,y) \to
\half (f(x) + f(y)) \,\delta (x-y)
\label{eq44}
\ee
can be taken.

To determine the subtraction, $c_F$, we fist compute in the Schr\"odinger
representation   the (regulated) unitary operator that implements the finite
transformation generated by $T_F$.
\be
U_F (\varphi_1,\,\varphi_2)
= \langle \varphi_1 | e^{- i T_F} | \varphi_2 \rangle
\label{eq45}
\ee
[As is well known, the exponential of an operator is better behaved than the
operator itself: $\langle \varphi_1 | T_F | \varphi_2 \rangle$
involves a functional $\delta$-function, $U_F (\varphi_1, \varphi_2)$
is an ordinary functional.]
Since $T_F$ is quadratic in the fields, $U_F$ can be explicitly constructed and
one can
explicitly study the limit (42).  It turns out that $U_F$ does not possess a
well defined limit, but
the singularities can be identified and isolated.  It is
found that
they are    contained    in a c-number    phase $e^{-i c_F}$, and $c_F$
is the required subtraction in the generator $T_F$.  Explicitly one finds
\be
c_F = - {1 \over 4\pi} \, P \int dx \, dy {F(x,y) \over (x-y)^2}
\label{eq46}
\ee
and this then has the consequence that
\be
\left[ \widetilde{T}_{f_1}, \, \widetilde{T}_{f_2} \right]
= i \, \widetilde{T}_{(f_1,f_2)} - {i \over 48 \pi}
\int dx \left( f_1 f_2''' - f_2 f_1 ''' \right)
\label{eq47}
\ee
where the tilda denotes well-defined,
subtracted generators and $(f_1, f_2)$ is the Lie
bracket of $(f_1,f_2)$.  The last term is of course the unavoidable Schwinger
term; it is a quantum anomaly in the commutator (13b,c) and changes the
constraints from first to second class.

Another perspective on the obstruction is given by the operator solution to
the theory, which may be explicitly constructed.  Returning to the formulation
in
(37--39), we see from (16--18)
that the classical solution is (with $\alpha=0$, so that
$\omega_1 = 0$)
\beq
\varphi &=& \phi_+ (X^+) + \phi_- (X^-) ~~,~~~~
\Pi = {d \over dx} \left( \phi_+ (X^+) - \phi_-(X^-) \right)
\nonumber\\
r^\pm &=&  Q^\pm (X^\pm) + {g\over2} N_\mp (X^\mp)
{} ~~,~~~~ \pi_\pm = \mp {1\over2} {d \over dx}
\left( N_\pm (X^\pm) - {2\over g} Q^\mp (X^\mp) \right)
\label{eq48}
\eeq
The quantities in (46) are promoted to operators, and the canonical
commutators of momenta with coordinates are reproduced provided we postulate
\beq
{}\left[ \phi_\pm (\xi), \, \phi_\pm(\xi') \right]
&=& -{i \over 4} \epsilon(\xi-\xi') \label{req48a}\\
{}\left[ Q^\pm (\xi), \, N_\pm (\xi') \right]
&=& {i \over 2} \epsilon(\xi-\xi') \label{req48b}
\eeq
with all other commutators vanishing.
[It is assumed that $X^+ (X^-)$ is an increasing (decreasing) function of $x$.]
The mode operators are chiral fields.
It   remains    to solve     the diffeomorphism constraints.   According   to
(19), this    would be achieved by enforcing the equality
\[
( \phi'_\pm )^2 = \half \left( N'_\pm Q'^\pm + Q'^\pm N'_\pm \right)
\]
at least weakly, {\it i.e.\/} when acting on states.  But such an equality
cannot hold: the left side --- the matter energy momentum density ---
possesses a Schwinger term in its commutator, the right side does not.

Gravity without matter as well as the semiclassical   limit  are obtained
\cite{ref11}
by expanding the phase of the wave functional in powers of the gravitational
constant $G$;
for us the expansion parameter will be $g \equiv 8\pi G / \lambda$:
$\Phi = \exp i\left(g^{-1}S_g + S_0 + gS_1 + \ldots \right)$.
We then operate on $\Phi$ with the energy constraint (36a) and equate to zero
powers of $g$.
The first few orders give the equations
\begin{mathletters}
\label{eq49}
\beq
g^{-2} ~&:&~~ {\delta S_g \over \delta \varphi} = 0
\label{eq49a}\\
g^{-1} ~&:&~~ {\delta S_g \over \delta r^a} \, {\delta S_g \over \delta r_a}
+ r'^a r'_a = 0
\label{eq49b}\\
g^0 ~&:&~~ \left( - {1\over2}
{\delta^2 \over \delta \varphi \delta \varphi}
+ {1\over 2}  \varphi'^{2} \right) \, e^{i S_0}
= -i \, {\delta S_g \over \delta r^a} {\delta e^{i S_0} \over \delta r_a}
- {i\over2} \, {\delta^2 S_g \over \delta r^a \delta r_a} \, e^{i S_0}
\label{eq49c}
\eeq
\end{mathletters}%
Eq.~(\ref{eq49a}) indicates that $S_g$ is
all that survives
when matter is absent and
is determined by Eq.~(\ref{eq49b}) to be
\be
\Phi(r^a)
\bigg|_{\hbox{\svrm pure gravity}}
= e^{{i\over2g} \int dr^a \epsilon_{a b} r^b}
{} ~~,~~~~ {\delta^2 S_g \over \delta r^a \delta r_a} = 0
\label{eq50}
\ee
which is also invariant against $x$-reparametrization --- as anticipated, the
momentum constraint is met automatically.  This explicitly shows the absence
of an obstruction --- the Schwinger terms cancel owing to the opposite signs
in the gravitational energy.        [When (\ref{eq50}) is used in (32) with
(34) and
the relation between $M$ and $m$ is recalled, we regain the pure gravity wave
functional (20).]    But the addition of     matter adds positive expressions
to the energy density, with non-canceling Schwinger terms, and a solution
cannot be found.\footnote
{\advance\baselineskip by -8pt
{}~If one has an even number of matter fields, $\varphi_n$, and is willing to
replace half of them by $i \, \varphi_n$, then a functional that is
annihilated pairwise by the matter energy densities can be constructed.  From
(50) one
sees that the functional involves
$e^{\pm {1\over2} \int \varphi_n d \varphi_{n+1}}$.
But this is a real exponential, rather than a phase, and diverges for large
fields.   We view such a ``solution'' to be unacceptable.  It was proposed in
Ref.~\cite{ref12}.}

The last topic that we address    concerns     a possible     semi-classical
 limit       of the quantum theory.  The
semi-classical approach consists of setting
\begin{mathletters}
\beq
\Phi &=& e^{{i\over2g} \int d r^a \epsilon_{ab} r^b}
\, \Phi_{\rm matter}
\label{eq51a}
\eeq
with $\Phi_{\rm matter}= e^{i(S_0 + g S_1 + \ldots)}$
satisfying equation (49c), which is exact as $g \to 0$,
\beq
\left( - {1\over2} {\delta^2 \over \delta \varphi \delta\varphi}
+ {1\over2} \varphi'^2 \right) \, \Phi_{\rm matter}
&=& r'^a \epsilon_a^{~b} \, {1\over i} \,
{\delta \Phi_{\rm matter} \over \delta r^b}
\eeq
\end{mathletters}%
Next one supposes that the $r^a$ dependence in $\Phi_{\rm matter}$ is such
that the right side of (51b) may be written as
$i {\delta \over \delta \tau(x)} \Phi_{\rm matter}$,
where $\tau(x)$   is an   emergent local time coordinate that allows viewing
(51b) as
a Schwinger-Tomonaga-type equation.   However here this is not possible, again
because of an obstruction:   ${\delta \over \delta \tau(x)}$
obviously commutes with itself
at all points on the line, but $r'^a(x) \epsilon_a^{~b} {\delta \over
\delta r^b (x)}$ does not.   Thus a sensible semi-classical description,
which still retains a limiting form of the diffeomorphism constraints, cannot
be given.

Evidently lineal gravity with matter is an anomalous theory.   We leave it for
future investigations to decide whether any of the familiar devices for
relaxing incompatible quantum constraints can be employed here to define an
acceptable lineal quantum gravity.   But even should this be possible, it is
doubtful that the semiclassical black hole puzzles
have a quantum analog in this model.

We are grateful for helpful conversations with M.~Ortiz.

\goodbreak

\end{document}